# LEARNING ANALYTICS DASHBOARDS FOR ADVISORS – A SYSTEMATIC LITERATURE REVIEW

Suchith Reddy Vemula and Marcia Moraes

Department of Computer Science, Colorado State University,
Fort Collins, Colorado

## ABSTRACT

*Learning Analytics Dashboard for Advisors is designed to provide data-driven insights and visualizations to support advisors in their decision-making regarding student academic progress, engagement, targeted support, and overall success. This study explores the current state of the art in learning analytics dashboards, focusing on specific requirements for advisors. By examining existing literature and case studies, this research investigates the key features and functionalities essential for an effective learning analytics dashboard tailored to advisor needs. This study also aims to provide a comprehensive understanding of the landscape of learning analytics dashboards for advisors, offering insights into the advancements, opportunities, and challenges in their development by synthesizing the current trends from a total of 21 research papers used for analysis. The findings will contribute to the design and implementation of new features in learning analytics dashboards that empower advisors to provide proactive and individualized support, ultimately fostering student retention and academic success.*

## 1. INTRODUCTION

Learning analytics (LA) is an intelligent use of data that helps educators and advisors identify behavioural and study patterns in learners by processing, reporting, and working on their data to optimize the learning environment [1]. By exploiting the already available data extracted from the activity of the students at the distance learning platforms, Learning Analytics Dashboards (LAD) significantly impact the learning and teaching processes for different stakeholders [2]. Learning analytics offers new opportunities to enrich feedback practices in higher education [3], which influences student motivation. Advisor dashboards provide feedback to students by looking at their progress to improve their self-regulated learning [4].

Academic advising is a decision-making process that assists students in the clarification of their career and life goals and the development of an educational plan for the realization of these goals through communication and information exchanges with an advisor [5]. The key features and functionalities required for an advisor-focused LAD are centred around the early identification of at-risk students, personalized support strategies, integration with various data sources, predictive analytics capabilities, clear data visualization, and adherence to privacy and ethical guidelines [6]. The field of learning analytics is continuing to progress from initial scenarios of using data to predict which students are at risk of withdrawal or academic failure to more generic scenarios in which data are used to decide how best to improve the learning experience and academic outcomes of all students. The phrase closing the loop has been used to describe the need to connect actions (or interventions) back to the learners. Yet, this connection has not been sufficiently explored [7]. The rationale for this paper in the context of existing knowledge tries to explore that connection and fill the gap from the perspective of the advisors, as there is not so much data and work available on the dashboard for advisors.

Considering this context, we conducted a literature review to understand the current state of the art in LAD for three different stakeholders: students, instructors, and advisors. As our interest relies on advisors, we then focused on analysing LAD for that specific kind of stakeholder. In this sense, our paper aims to answer the following research questions:

**RQ-1:** What is the current state of the art related to LAD considering different stakeholders, such as students, instructors, and advisors?

**RQ-2:** What key features and functionalities were identified in a LAD for advisors to provide better advising options to students?

The remainder of this paper is structured as follows. Section 2 presents the methodology applied in this study. Section 3 will present the results. We discuss the results in Section 4 and outline directions for further work and provide our concluding remarks in Section 5.

## 2. METHODOLOGY

The selection of relevant research papers for a literature review requires a systematic approach based on well-defined inclusion and exclusion criteria. In this review, we establish our inclusion criteria to encompass studies published in peer-reviewed journals or conference proceedings within the last five years (2018–2023) that examine the learning analytics domain and learning analytics dashboards. We also sought research that explored the previous literature review on LA Dashboards for perspectives and experiences, as well as the impact and effectiveness of these dashboards in the present scenario, to get an understanding of what has already been implemented. On the other hand, we excluded non-peer-reviewed articles, publications prior to 2018, and duplicate or redundant studies.

To conduct our search, we utilized multiple databases, such as Google Scholar, the ACM Digital Library, the Wiley Online Library, and the IEEE Library. These databases were selected because they are known for their publishing works in learning analytics. We employed a hierarchical search strategy, starting with general keywords and gradually narrowing down to more specific ones. Each database was subjected to filtering techniques, considering keywords like "Learning Analytics," "Dashboards," "Advisors," "Feedback-based learning," "Problem-Based Learning," and "Self-Regulated Learning."

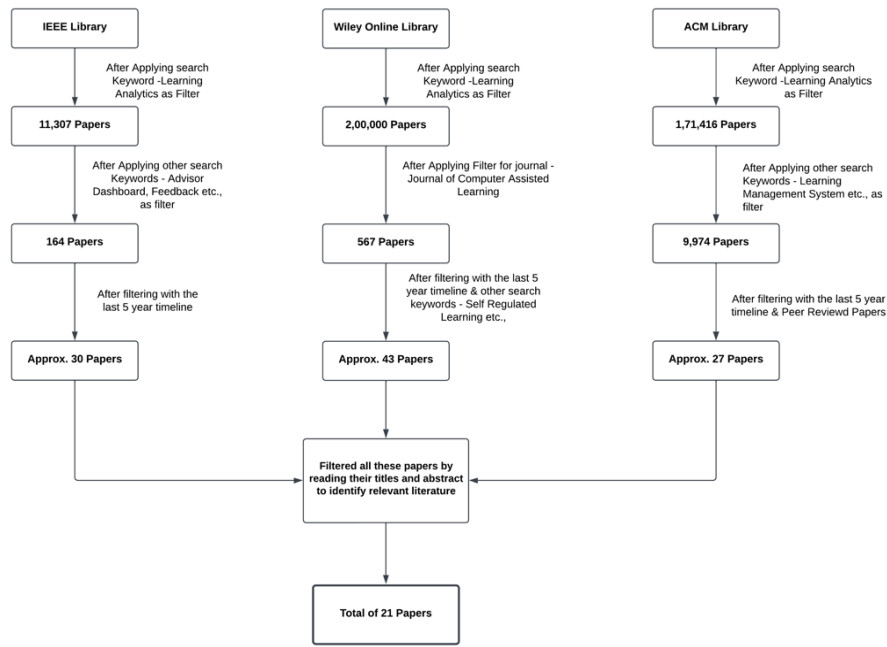

Figure 1. Selection Process of Published Works for Literature Review

As per the Figure 1., in the IEEE database, we initially retrieved 11,307 papers using the "Learning Analytics" filter. Further application of filters such as "dashboard," "advisors," and "self-regulated learning" resulted in the selection of 164 relevant papers. In the case of the Wiley Online Library Database, we filtered out over 2,00,000 papers by narrowing down the search to the past five years and focusing on the Journal of Computer-Assisted Learning, which yielded 567 papers relevant to feedback-based learning and dashboards for advisors. Similarly, the ACM Digital Library search produced 1,71,416 results for learning analytics dashboards, and subsequent filters led to 9,974 papers relevant to advisors, self-regulated learning, and feedback-based learning. Google Scholar was used as a supplementary research tool to help us make our literature review more comprehensive. We used keywords and phrases related to our study topic in systematic searches to get the most out of it. Google Scholar's extensive database allowed us to identify additional relevant journals and papers that may not have been included in our initial search. We then looked over and evaluated these new sources to see if they fit with our research and what they could add. This made sure that our literature review was as comprehensive and up to date as possible.

To ensure the eligibility of studies for synthesis, we followed a systematic process. This involved formulating inclusion criteria aligned with our research question and conducting a thorough screening and selection of studies. Initially, we reviewed titles and abstracts to isolate papers that matched our research focus. Subsequently, the selected papers underwent full-text evaluation to assess methodology, research questions, data quality, and analysis. We also maintained research integrity by conducting critical appraisals to assess credibility, robustness, and potential bias. Relevant data was then extracted from the final set of included studies, and we tabulated intervention characteristics. We compared these characteristics against different stakeholders, including students, instructors, and advisors, to ensure relevance and address specific research objectives. Notably, we chose manual screening over automation tools to assess the alignment of papers with our research objectives and select them accordingly. Overall, this systematic approach allowed us to compile a comprehensive collection of research papers addressing the current state of learning analytics dashboards for students, instructors, and advisors and made it possible for us to analyse more in depth about the works done focusing on the advisors, and the impact of LAD on student outcomes and advisor performance.

## 3. RESULTS

Following the systematic review process outlined above, we identified and selected a total of 21 papers for inclusion in our study. Within this selection, 3 papers are previous literature reviews, while the remaining 18 papers focus on Learning Analytics Dashboards (LA Dashboards). Specifically, among these 18 papers, 7 are oriented towards students, 6 are geared towards instructors, and 5 are designed for advisors. In the forthcoming sections, we will provide concise descriptions of these selected papers to offer a comprehensive overview of our research findings.

### 3.1. Previous Literature Review on LA and LAD

In prior literature reviews on LA and LAD, three papers have further enriched our understanding of this field. A literature survey conducted by Mittal et al. [8] takes a qualitative descriptive approach to delve into learning analytics within higher education. The study provides a clear definition of learning analytics, outlines its cyclic nature, and explores the challenges and benefits associated with its implementation. Moreover, it underlines the intersection between LA and Educational Data Mining (EDM), emphasizing the role of user behaviour in the learning process. The paper further explores the stages of learning analytics, encompassing prediction, clustering, relationship mining, model discovery, and human resource data separation. It culminates by highlighting the advantages of LA, including course relevance identification, infrastructure enhancement, personalized learning opportunities, and post-learning prospects. However, it also recognizes limitations of online learning, particularly in contexts involving hands-on experiential learning.

Timmi et al. [1] reviewed 24 papers as a part of their literature survey and delved into LA, describing its role, goals, and benefits. It defines LA as an intelligent use of data to optimize learning environments by analysing learner's behavioural and study patterns. LA's multifaceted goals encompass enhancing student and teacher performance, improving educational content, aiding slower learners, precise assessment, supporting instructor development, and optimizing resource allocation. Various tools like DataShop, GISMO, Inspire, LOCO-Analyst, Meerkat-ED, MDM Tool, Performance Plus, SNAPP, and Solutionpath StREAM are discussed for research in LA. The paper elaborates on methods employed, such as causal mining, clustering, discovery with models, knowledge tracing, outlier detection, prediction, recommendation, and more. It concludes by highlighting the beneficiaries of LA who are students, instructors and academic authorities mentioning what kind of benefits they get from LA and emphasizing an understanding of LA's objectives, tools, and methods.

In their systematic literature review, Banihashem et al. [3] contributed to our understanding of LA and LAD by focusing on how LA can be used to improve feedback practices in higher education settings where technology is used. In doing so, the paper highlights the benefits of LA in providing timely and constructive feedback to students, aiding their learning progress and self-regulation. It also sheds light on understanding how LA tools can effectively support feedback procedures in higher education. To provide a comprehensive view, the authors conducted a rigorous literature review using the PRISMA methodology, considering 46 relevant publications out of an initial pool of 1318. Through this analysis, the study seeks to review and answer these four main dimensions of the analytical framework: why LA is used for feedback, what types of data LA employs, which methods it uses, and how LA benefits different stakeholders (educators and students). Their findings reveal that LA-supported feedback serves various purposes, including reflection, personalization, prediction, assessment, adaptation, and recommendation. The types of data employed range from trace data to assessment data and demographic data. Analytical methods encompass information visualization, data mining, text analytics, and social network analysis. As a culmination, the review provides a conceptual framework to guide the effective use of LA for feedback and strongly advocates for future research that investigates the impact of different data types on feedback practices within higher education.

It is important to note that none of these previous literature reviews specifically focused on examining LAD for advisors. This represents an important gap in the previous and existing literature, as LADs play a critical role in supporting advisors in their interactions with students and decision-making processes. Further research and exploration in this area could provide valuable insights into how LADs can be tailored to meet the specific needs and objectives of advisors in higher education settings.

## 3.2. Papers on LA Dashboards for Students

Out of the 19 papers on LAD, seven focused on LAD designed to assist students in enhancing their learning experiences. Through the analysis, it becomes evident that these papers centred around feedback-based learning and self-regulatory learning.

A significant contribution to the discussion comes from Bodily et al. [9], who introduce a student-centred learning analytics dashboard (LAD) with content and skill recommenders for improved online learning. Their work addresses not only the technical aspects, including the use of xAPI and open-source Learning Record Stores (LRS), but also delves into dashboard design, functionality, and its reception among students. The content recommender aims to improve mastery through iterative design, while the skilled recommender focuses on nurturing metacognitive strategies, employing an innovative radar chart approach. Importantly, the study incorporates student feedback, emphasizing the pivotal role of student engagement and user-friendly design in shaping effective learning analytics dashboards.

Lim et al. [10] investigate how students make sense of learning analytics-based personalized feedback and adapt their self-regulated learning (SRL) processes accordingly across four courses. The paper underscores the importance of feedback in enhancing students' learning progress and examines its influence on their self-regulation. The authors propose several implications for practice, including aligning personalized feedback with individual learning preferences, maintaining coherence with course syllabi, adopting a positive tone, and enhancing dialogic communication. The study employs a dual framework involving student perceptions of feedback and the sociocognitive SRL model. The research addresses two primary research questions concerning students' sense-making of personalized feedback and variations across different contexts. The investigation uses software called OnTask for delivering personalized feedback to students and employs epistemic network analysis (ENA) to explore associations between students' feedback perceptions and their SRL adaptations. The study highlights students' enhanced motivation, reduced procrastination, and improved goal setting resulting from personalized feedback, demonstrating the interplay between feedback and SRL processes. Limitations of the study are acknowledged, such as the reliance on self-reported data and the omission of learner characteristics, which could be addressed in future research. The study contributes to understanding the impact of learning analytics-based personalized feedback on students' learning experiences.

The study by Yoo and Jin [11] focuses on identifying challenges in online discussion activities and creating visual design guidelines to address them. The research employs a prototyping methodology to develop five prototype dashboards for learning analytics in online discussions.

These dashboards provide information on participation, interaction, and discussion content keywords, message types, and debate opinions. The development process includes expert validation, usability testing, and user experience evaluations. The goal is to enhance learners' self-regulation by offering visual feedback on their participation and interactions in online discussions. The study emphasizes the need for process-oriented feedback and explores the perceptions of learners and instructors towards the developed dashboards. However, the research lacks implementation within an actual online learning system and does not consider interventions for individual differences.

Iraj et al. [12] focused on embedding trackable Call to Action (CTA) links in feedback messages to enhance students' engagement with the feedback and improve course success prediction. Feedback effectiveness hinges on students understanding, acting upon, and benefiting from it. The paper emphasizes feedback literacy and distinguishes between different levels of feedback: task, process, and self-regulatory. The authors address the challenge of the "feedback gap" and explore technology-supported feedback systems. They propose using CTA links in personalized feedback messages to monitor engagement. The research questions examine the association between engagement and success, demographic factors influencing engagement, and students' perceptions of actionable feedback. The study uses logistic regressions and a focus group interview to answer these questions.

Wang et al. [13] explored the effectiveness of Learning Analytics Dashboards (LADs) based on process-oriented feedback in the context of iTutor, an e-learning platform, in their paper. The study conducts an experiment with two groups: one using LADs and the other using Original Analytics Reports (OARs) with product-oriented feedback. The research aims to validate two hypotheses: 1) Process-oriented feedback is more effective for student learning compared to product-oriented feedback, and 2) LADs benefit students with low prior knowledge more than those with high prior knowledge. The OAR in iTutor provides static explanations of learners' performance, while LADs offer interactive, personalized, and analytical feedback through various formats like statistics, text, and graphics. The study finds that LADs are more effective in enhancing e-learning compared to OARs. LADs particularly benefit students with low prior knowledge, making their learning outcomes like those of those with higher prior knowledge.

Lim et al. [14] investigated the impact of feedback on students' self-regulated learning (SRL) processes in different course contexts. It explores how feedback influences learning tactics, strategies, and time management. The research uses mixed methods, analysing learning traces and students' recall of experiences. The study identifies distinct learning tactics and strategies, categorizing students into three engagement patterns: Strategic-Moderate Engagement, Intensive-High Engagement, and Highly Selective-Low Engagement. The findings suggest that personalized feedback positively influences engagement and performance. Additionally, students' comments were analysed to understand SRL adaptations in response to feedback. However, the study acknowledges limitations in not considering demographic and individual learning characteristics.

Duan et al. [4] conducted a study that aimed to improve students' self-regulated learning and course performance through a designed LAD by exploring its impact using mixed methods. The dashboard's goals included raising awareness of the link between learning behaviour and progress and motivating effective learning strategies. Research questions focused on influential learning activity features, LAD design, and its impact. The investigation was conducted in a Python course, involving student log data to identify impactful learning behaviours. The LAD was iteratively designed, tested on students, and evaluated through surveys and interviews. Limitations included a small sample size from a single business course, potentially affecting generalizability, and the non-random assignment of dashboard users, introducing bias.

### 3.3. Papers on LA Dashboards for Instructors

We analysed six papers related to LAD for instructors. One of those papers was written by Pardo et al. [7] centred around the OnTask tool, which integrate personalized learning support actions (PLSAs) within a Learning Management System (LMS). The paper presents three main contributions: 1) A student-instructor-centred conceptual model connecting student information with instructor-created rules for PLSAs; 2) a software architecture comprising six functional blocks for deploying PLSAs; and 3) the implementation of this architecture as an open-source platform named OnTask. The model and architecture aim to bridge the gap between algorithmic solutions and practical instructor-learner interactions by providing a mechanism for instructors to design PLSAs based on data insights. The model encompasses six phases, including data warehousing, data import, student data table creation, creation of PLSAs, integration of data mining and machine learning algorithms, and implementation of PLSAs. The architectural design involves components for authentication, data management, SDT and PLSA creation, and a notification gateway REST API. The implemented platform enables instructors to specify and deploy scalable, data-supported personalized student support processes, enhancing the learning experience through personalized feedback and actions.

In the domain of learning analytics, several studies have contributed to the ongoing discourse by leveraging the power of machine learning and data analysis. Aljohani et al. [15] research introduces the AMBA Prototype, a learning analytics framework for managing learning data aimed at optimizing decision-making in education. This tool seamlessly integrates a LAD with LMS, providing instructors with valuable insights. Notably, Natural Language Processing (NLP) techniques and BERT models are harnessed to analyse students' engagement with the Blackboard System, yielding significant improvements in student performance and engagement, as evidenced by an MANOVA test results. These studies collectively advance the field of learning analytics by harnessing machine learning and cognitive state analysis and offering innovative solutions for educators to enhance student learning experiences.

Furthermore, Zilong et al. [16] tackle real-time interventions in problem-based learning (PBL) by proposing machine learning-driven solutions for processing and presenting user-generated data in a dashboard. Their research focuses on text data processing and problem-solving tracking. A case study involving an Alien rescue PBL task demonstrates the use of NLP and BERT models for automated text analysis. Flow theory-based sequential analysis reveals cognitive states during problem-solving. The study underscores BERT's effectiveness in handling extensive textual data, aiding teachers in identifying students needing support. By employing descriptors like flow, anxiety, and boredom, the research enables effective tracking of student progress and interventions.

Shibani et al. [17] presents an in-depth investigation into educators' perspectives on the implementation of learning analytics, focusing on a writing feedback tool called 'AcaWriter'. Through qualitative analysis of interviews with lead educators, the study delves into motivations driving educators to adopt learning analytics, strategies they employ for successful integration, challenges encountered, and observed outcomes. Findings reveal educators' motivations to enhance communication skills, provide formative feedback, save time, and embrace technology's potential. Implementation strategies for 'AcaWriter' include co-design, designing authentic experiences, receiving support, and fostering wider adoption. Challenges for implementation involve addressing tool limitations, guiding students in utilizing automated feedback effectively, and managing increased effort. Outcomes highlight improvements in self-assessment, feedback provision, time efficiency, enhanced instructor knowledge, and contributions to writing research. Tsai et al. [18] have reviewed a pilot study about OnTask in which the study examines students' perceptions of the feedback they receive via the OnTask system while considering their self-efficacy and self-regulation abilities. OnTask integrates personalized feedback using rules based on "if this, then that" statements, allowing instructors to send tailored messages to students based on course-related parameters. The study addresses three research questions: 1) whether students' overall feedback experience improved after implementing OnTask; 2) how students' self-efficacy and self-regulation relate to their feedback perception; and 3) the gaps between the perceived importance of feedback and students' OnTask-generated feedback experience. The methodology involves conducting surveys before and after introducing OnTask and performing exploratory data analysis. The results reveal that OnTask-enhanced feedback positively impacted students' relational experiences, particularly two affect-level aspects which are "relationship with the teacher" and "self-confidence". Students with high self-efficacy and self-regulation exhibited more positive feedback experiences. However, OnTask's performance fell short in aiding domain knowledge development and self-regulated learning. The study suggests that both instructor feedback literacy and the standards for generating feedback in OnTask need further investigation. Nonetheless, this pilot study highlights the potential of OnTask to facilitate a dialogue between teachers and students while emphasizing the role of student dispositions in feedback engagement. The paper by Dourado et al. [21] outlines the development and initial evaluation of a Learning Analytics Dashboard (LAD) designed to furnish online instructors with process-oriented feedback. The research involved an iterative human-centred design process, encompassing two rounds of understand, ideate, and make activities. In the first iteration, ten teachers were interviewed to comprehend their needs and practices, and nine teachers took part in co-design sessions using paper prototypes. Subsequently, five visualization experts conducted a heuristic evaluation of the dashboard in the second iteration. The findings revealed data and task requirements for process-oriented teacher feedback and introduced a dashboard with coordinated views for overall, pattern, and detail analysis. While the expert reviews were generally positive, issues arose concerning the timeline and retrospective pattern view. The paper concludes by expanding prior data and task taxonomies for process-oriented feedback, offering preliminary evidence of dashboard usability, but acknowledging the need for end-user validation. It reflects on design challenges related to data cleaning, time representation, and visualizing "good" versus "bad" paths and calls for future work to address expert review issues and validate with teachers, thereby contributing an empirically grounded dashboard design to inform future Learning Analytics Dashboard research focused on process-oriented feedback for teachers.

## 3.4. Papers on LA Dashboards for Advisors

Only five papers on LAD's designed specifically for advisors were identified. Among them, Gutiérrez et al. [5] presented the design and concept of a Learning Analytics Dashboard for Advisers (LADA) aimed at enhancing academic advising through comparative and predictive analysis. While existing research has focused on the benefits of learning analytics for students and instructors, this study highlights the potential benefits for academic advisors. LADA aims to support advisors in making informed decisions by providing detailed insights and prediction models. LADA compiles a semester plan for students based on their academic history, predicts academic risk using clustering techniques, and offers a user-friendly interface. LADA's interface consists of two sections: one displaying the chance of success for students in a course and the other providing various components in the form of information cards. The prediction of success is based on multilevel clustering techniques using historical data and current course selection. The study conducted user studies at two different universities the undergraduate level to assess LADA's utility and usability. In University A, the study recruited six employees from the academic advising service who regularly advised on semester planning for undergraduate students, and six PhD students with no advising experience. In University B, the study recruited five academic advisers who regularly advised on semester planning for undergraduate students. Therefore, a total of 17 advisers participated in the study. Results showed that participants found LADA appealing but suggested the need for more transparency in the algorithm's predictions to increase confidence. However, LADA enabled advisors to evaluate a greater number of scenarios in a similar amount of time, particularly for challenging cases, enhancing the decision-making process. Overall, LADA is perceived as a valuable tool for both experienced advisors and laymen, offering accurate decision-making support.

Additionally, Charleer et al. [19] have also made significant contributions by focusing on the creation, development, and evaluation of a Learning Analytics Dashboard for Insights and Support During Study Advice (LISSA) to enhance communication between study advisors and students. The dashboard aims to provide a clear overview of student study progress, encourage peer comparison, and facilitate data-driven discussions between advisers and students. The iterative design process and evaluation findings from 97 counselling sessions are reported. Unlike other dashboards, LISSA is tailored to meet the needs of study advisers, offering personalization, factual insights, and interactive features to guide advising sessions. The contribution of the paper lies in presenting the dashboard's design journey, evaluating its usability and utility for advisers, analysing the impact of dashboard-supported sessions, and providing guidelines for similar dashboard developments. The requirements were gathered through observation and brainstorming sessions with study advisers. The dashboard was evaluated post-exams and found to facilitate dialogue, motivate students, and trigger meaningful discussions. However, further research is needed to ascertain its applicability to different programs and institutions with varying data sizes and student types.

Hilliger et al. [6] explored the implementation of academic advising dashboards in Latin American universities to enhance student decision-making and academic progress. It outlines the development of two types of dashboards: one relying on descriptive analytics to inform advisors about student interactions with the learning management system, and another using predictive analytics, driven by machine learning, to identify at-risk students and offer targeted interventions. The paper acknowledges the challenges of creating effective advisory dashboards, including the need for relevant indicators and the differing perspectives of stakeholders. It emphasizes that many existing dashboards prioritize staff needs over students' and highlights a study conducted at four Latin American universities (U1–U4) to gather insights on advising practices and experiences. Notable examples from the literature include the University of Michigan's Early Warning System (EWS) and Georgia State University's GPS advising system, which employ predictive analytics to provide interventions for students at risk. The authors advocate for future research to focus on co-designing student-oriented dashboards and integrating both descriptive and predictive analytics indicators for comprehensive support.

Another paper by Scheers and De Laet [20], which introduces exploratory research concerning the design and evaluation of a dashboard that integrates a black-box predictive model for student success into advising practices for university students. The authors examined existing literature, revealing a need for accurate and explainable predictive models. They found that most previous studies focused on program-level predictions, often relying on less powerful, explainable models. The research aims to incorporate predictive models into a Learning Analytics Dashboard (LAD) for advising and enhancing data-based support. However, the challenge lies in the black-box nature of these models, as they lack transparency and can undermine user trust and interpretability. The authors propose an interactive and explainable dashboard to mitigate this issue, facilitating insights and reflection while addressing legal, financial, and ethical concerns associated with predictive model adoption in higher education.

De Laet et al, [22] presents a comprehensive case study on the adoption and impact of a Learning Analytics Dashboard (LAD) designed for academic advising within a Latin American higher education institution. The study took place at the Escuela Superior Politecnica del Litoral (ESPOL) in Guayaquil, Ecuador, with the participant sample comprising student advisors and students who received academic advising through the new LAD. Research methods included thematic analysis of advising dialogues, surveys of student advisors, and surveys of students benefiting from LAD-guided advising. The LAD introduced new modules intended to enhance the dialogue between advisors and students, especially during the planning stage and discussions about academic history. The findings revealed that these new modules were predominantly utilized when advisors discussed specific subjects, academic performance, difficulties, and workload. Importantly, the study demonstrated that these new modules positively impacted perceived support, as student advisors reported better comprehension of students' needs and the ability to provide more personalized advice. The study concludes by emphasizing the importance of aligning evaluation goals with the LAD's intended purposes, considering the implementation context, engaging stakeholders in design, and providing training and support to ensure effective LAD utilization. Overall, this case study provided valuable insights into the successful adoption and impact of LADs for academic advising in higher education.

## 4. DISCUSSION

Learning Analytics Dashboards (LADs) offer valuable insights and support in education. This Literature review examined 19 papers that involved three different stakeholders: students, educators, and advisors, and addressed the proposed research questions as follows.

**RQ-1:** What is the current state of the art related to Learning Analytics Dashboards considering different stakeholders, such as instructors, advisors, and students?

A significant emphasis on personalized feedback-based learning and self-regulated learning (SRL) is present in the current landscape of LAD for students. LA Dashboards aim to empower students with timely and actionable feedback to enhance their educational journey, aligning with the broader trend of recognizing the value of guidance in improving student outcomes. SRL, a central theme, underscores the importance of enabling students to take charge of their learning processes, fostering metacognitive skills crucial for lifelong learning [10]. User-centred design principles, usability, and student engagement play a pivotal role in shaping effective LAD, acknowledging that their impact hinges on student adoption. The diverse approaches to LAD, including content recommenders, process-oriented feedback, and visual design guidelines, reflect an evolving field that combines pedagogical and technical expertise. While empirical effectiveness studies are common, challenges such as the "feedback gap," potential biases, and the need for larger-scale research are recognized.

The current state of LAD for instructors emphasizes personalized learning support actions (PLSAs) and data-driven decision-making. Tools like OnTask integrate PLSAs within LMS, enabling instructors to design scalable, data-supported personalized student support processes [7]. The integration of machine learning, natural language processing (NLP), and cognitive state analysis enhances instructors' abilities to analyse student engagement and optimize decision-making [15]. Real-time interventions through machine learning-driven solutions help instructors track student progress effectively. Educators are motivated to adopt learning analytics tools for enhancing communication, providing formative feedback, and saving time. However, further exploration of instructor feedback literacy and standards for feedback generation is needed to maximize the potential of these dashboards in education.

The existing landscape of LAD for academic advisors showcases their potential to revolutionize academic advising. Notable examples like LADA and LISSA offer advisors detailed insights and predictive models to facilitate data-driven discussions and informed decision-making. LADA, for instance, compiles semester plans and predicts academic risk, providing advisors with valuable tools [5]. While advisors find these dashboards appealing, there's a need for greater transparency in prediction algorithms to enhance confidence. Academic advising dashboards are also not confined to specific regions, with Latin American universities adopting similar systems that incorporate descriptive and predictive analytics, emphasizing the importance of predictive models in student success support [6]. However, a challenge arises from the black-box nature of predictive models, which lack transparency and pose concerns about user trust. To address this, research is exploring interactive and explainable dashboard designs, focusing on insights and reflection while addressing legal and ethical considerations associated with predictive models in higher education. This highlights a growing emphasis on making LAD user-friendly and transparent for advisors, aligning with evolving academic advising needs.

**RQ-2:** What key features and functionalities were identified in a learning analytics dashboard for advisors to provide better advising options to students?

In our current exploration, we delve into the key features and functionalities identified in Learning Analytics Dashboards (LADs) designed to empower advisors in providing effective support and data-informed decisions to students. These dedicated dashboards, tailored for advisors, significantly enhance the academic advising process through a range of essential features. Notably, predictive analytics take centre stage, enabling the identification of at-risk students and the timely initiation of interventions. These dashboards provide comprehensive student profiles, offer valuable intervention recommendations, and equip advisors with communication tools to foster meaningful interactions. Usability, characterized by a user-friendly design, and transparency in predictive algorithms are considered fundamental aspects of these tools. Personalization features empower advisors to tailor guidance to the unique needs of individual students, while feedback mechanisms facilitate seamless interaction. Furthermore, integration with Learning Management Systems ensures the synchronization of data, and scalability is a key consideration to accommodate growing student populations. Beyond these general features, specific research papers highlight innovative approaches in the field. For instance, LISSA promotes data-driven discussions between advisors and students [19], while Latin American universities draw inspiration from models such as the Early Warning System (EWS) and GPS advising to explore academic advising dashboards that employ descriptive and predictive analytics. Additionally, research underscores the importance of interactive and explainable dashboards to enhance user trust and address the legal and ethical concerns associated with predictive models in higher education. Collectively, these findings shape the evolving landscape of Learning Analytics Dashboards for advisors.

## 5. CONCLUSION & FUTURE WORKS

Building upon the foundational features and functionalities of the LAD discussed earlier, there is a recognized need for additional elements that can further empower advisors to assist students effectively. To fulfil this need, we believe it is crucial to align our system with the goals of the Early Performance Feedback Initiative. By incorporating this initiative into our LAD, we propose to integrate feedback and assessment data from the first four weeks of student learning. This addition acknowledges the importance of early performance data, reinforcing the idea that early effort and performance indeed matter. Furthermore, it directly supports the purpose of the Early Performance Feedback program by ensuring that advisors have immediate access to actionable insights into students' performance during the crucial initial weeks from the very beginning of the semester. In addition to the EPF, we recognize the need for another crucial element within our LAD. This essential feature involves the integration of cumulative, individual, and mean percentage analysis of student scores at various points during the semester. For instance, we propose implementing this analysis at each examination point that courses can have throughout the semester. This approach allows advisors to have a comprehensive historical view of student performance, enabling them to track progress over time. With this comprehensive tool at their disposal, advisors can assess and support students effectively in multiple dimensions. Advisors will not only have insights into students' early performance but also gain a holistic understanding of their academic progress. This integration empowers advisors to identify areas where students excel and where they may be encountering challenges, allowing them to proactively connect students to relevant academic support resources. By combining both these features, our LAD ensures advisors have the necessary tools to guide students toward academic success. These elements have been identified as valuable tools to enhance the advising process, yet they have not been extensively addressed in previous literature reviews or research papers.

To address this gap, surveys and interviews were conducted with Undergraduate Academic Success Coordinators to understand their expectations regarding these new features. And we gathered valuable insights from them which provided us with a comprehensive understanding of the specific needs and expectations surrounding these new features. Their input confirmed the significance of integrating the above discussed features into our Advisor Dashboard and highlighted the potential impact of incorporating them. In line with our commitment to addressing these needs and expectations, we aim to build a dashboard that incorporates these elements aligning with the evolving landscape of LA Dashboards for advisors and ensuring advisors have access to actionable insights for better student support and guidance with our further work in continuity with this Literature review.

While acknowledging the value of current features such as warning systems and predictive analysis, we recognize that there is room for substantial improvement in the advising process. With the proposed implementation of the features discussed above, we believe we can achieve a holistic and closed-loop system of Advising and Feedback, effectively addressing all gaps in the process of advising students through LADs. Our comprehensive literature review played a pivotal role in leading us to this conclusion. Therefore, we put forth the proposal to design an advanced LAD capable of seamlessly integrating the above discussed features which will not only facilitate advisors in enhancing their decision-making process but will also empower them to provide students with more personalized and effective support.

## AUTHORS

**Suchith Reddy Vemula** is a graduate student pursuing a Master's degree in the Computer Science Department at Colorado State University. He has a background as a Software Engineer Intern at DISH Network and as a Graduate Research Assistant under Marcia Moraes, where he focuses on Learning Analytics Dashboards. His research interests span various areas, including Computer Science Education, Learning Analytics, and Software Engineering & Development.

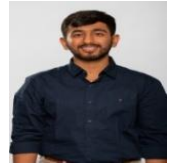

**Marcia Moraes** is an Assistant Professor at Colorado State University. She did a Post-Doctoral Fellow at the Center for the Analytics of Learning and Teaching (C-ALT), School of Education, Colorado State University. Her research interests are in learning analytics, epistemic network analysis, technology enhanced teaching and learning, computer science education, and distance learning. She is particularly interested in developing innovative applications and methods that improve learning and teaching.

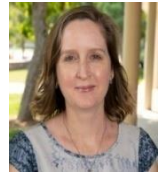